\documentclass[twocolumn,prl,showpacs,amsfonts,amsmath,assymb,eufrak]{revtex4}

\usepackage{epsfig}
\usepackage{color}
\usepackage{bm}

\usepackage{braket}

\usepackage{graphicx}

\newcommand{\eq}[1]{\begin{equation} #1 \end{equation}}
\newcommand{\eqa}[2]{\begin{equation} #1 \label{#2} \end{equation}}
\newcommand{\balign}[1]{\begin{align} #1 \end{align}}

\newcommand{\figin}[4]
{\begin{figure}[tb]
\centering
\includegraphics[width= #1]{#2.pdf}
\caption{#3}
\label{f:#4}
\end{figure}}

\newcommand{\todayd}{\the\year/\the\month/\the\day}

\newcommand{\bib}{\bibitem}

\newcommand{\lb}{\label}
\newcommand{\nt}{\notag}

\newcommand{\bel}{\begin{easylist}}
\newcommand{\eel}{\end{easylist}}

\newcommand{\eref}[1]{Eq.~\eqref{#1}}

\newcommand{\fref}[1]{Fig.~\ref{f:#1}}

\def \({\left(}
\def \){\right)}
\newcommand{\la}{\left\langle}
\newcommand{\ra}{\right\rangle}
\newcommand{\abs}[1]{\left|#1\right|}

\newcommand{\sumtwo}[2]%
{\mathop{\sum_{#1}}_{#2}}
\newcommand{\sumthree}[3]%
{\mathop{\mathop{\sum_{#1}}_{#2}}_{#3}}
\newcommand{\sumfour}[4]%
{\mathop{\mathop{\mathop{\sum_{#1}}_{#2}}_{#3}}_{#4}} 
\newcommand{\prodtwo}[2]%
{\mathop{\prod_{#1}}_{#2}}
\newcommand{\mintwo}[2]%
{\mathop{\min_{#1}}_{#2}}
\newcommand{\maxtwo}[2]%
{\mathop{\max_{#1}}_{#2}}
\newcommand{\maxthree}[3]%
{\mathop{\mathop{\max_{#1}}_{#2}}_{#3}}
\newcommand{\limtwo}[2]%
{\mathop{\lim_{#1}}_{#2}}
\newcommand{\suptwo}[2]%
{\mathop{\sup_{#1}}_{#2}}
\newcommand{\supthree}[3]%
{\mathop{\mathop{\sup_{#1}}_{#2}}_{#3}}
\newcommand{\supfour}[4]%
{\mathop{\mathop{\mathop{\sup_{#1}}_{#2}}_{#3}}_{#4}} 
\newcommand{\inftwo}[2]%
{\mathop{\inf_{#1}}_{#2}}
\newcommand{\infthree}[3]%
{\mathop{\mathop{\inf_{#1}}_{#2}}_{#3}}
\newcommand{\inffour}[4]%
{\mathop{\mathop{\mathop{\inf_{#1}}_{#2}}_{#3}}_{#4}} 

\newcommand{\bsp}{\boldsymbol{p}}

\newcommand{\para}[1]{{\em #1}\/.---}
\newcommand{\qex}{Q^{\rm ex}}
\newcommand{\pss}{p^{\rm ss}}
\newcommand{\dsgmhs}{\dot{\Sigma}^{\rm HS}}
\newcommand{\tw}{\tilde{W}}
\newcommand{\dsgm}{\dot{\Sigma}}

\def\rnum#1{\resizebox{0.5em}{\height}{\expandafter{\romannumeral #1}}}
\def\Rnum#1{\resizebox{0.5em}{\height}{\uppercase\expandafter{\romannumeral #1}}}

\begin{document}

\preprint{APS/123-QED}

\title{Speed Limit for Classical Stochastic Processes}

\author{Naoto Shiraishi}
\affiliation{Department of Physics, Keio university, Hiyoshi 3-14-1, Kohoku-ku, Yokohama, Japan}%

\author{Ken Funo}
\affiliation{School of Physics, Peking University, Beijing 100871, China}%

\author{Keiji Saito}
\affiliation{Department of Physics, Keio university, Hiyoshi 3-14-1, Kohoku-ku, Yokohama, Japan }%

\date{\today}

\begin{abstract}
Speed limit for classical stochastic Markov processes with discrete states is studied.
We find that a trade-off inequality exists between the speed of the state transformation 
and the entropy production.
The dynamical activity determines the time scale and plays a crucial role in the inequality.
For systems with stationary current, a similar trade-off inequality with the Hatano-Sasa entropy production gives a much better bound on the speed of the state transformation.
Our inequalities contain only physically well-defined quantities, and thus the physical picture of these inequalities is clear.

\end{abstract}


\maketitle

\para{Introduction}
Obtaining a fundamental bound on the speed of the state transformation is an important question relevant to broad research fields including quantum control theory, adaptation, and foundations of nonequilibrium statistical mechanics.
This question has been first investigated in an isolated quantum system, for which the derived bounds are nowadays called quantum speed limits~\cite{MT, Fle, AA, Pre, ML, GLM, JK,DC}. 
Consider a state transformation from a given initial state $\rho=\rho(0)$ to a given final state $\rho'=\rho(\tau)$ in a time interval $0\leq t\leq \tau$. 
Quantum speed limits claim that there is a trade-off relation between the operation time $\tau$ and the energy fluctuation. 
For instance, Mandelstam and Tamm derived the following expression 
\begin{align}
\frac{{\cal L} (\rho, \rho' )}{\Delta E_{\tau} / \hbar} \le \tau  \, , 
\nonumber
\end{align}
where ${\cal L} (\rho, \rho' )$ is the distance between the states $\rho$ and $\rho'$, and $\Delta E_{\tau}$ and $\hbar$ are respectively the energy fluctuation and the Planck constant~\cite{MT}.  This explicitly  shows the bound on the operation time $\tau$ in quantum state transformations.
Later, Margolus and Levitin~\cite{ML} derived another bound, where the bound on the operation time is characterized by the difference between the mean energy and the ground state energy. 
Generalization of the quantum speed limits to the case of shortcuts to adiabaticity \cite{Funo} and open quantum systems \cite{Tad, Cam, DL, Zha,Pir} have also been studied intensively.

The quantum speed limit has its origin in the uncertainty relation between time and energy.  Indeed, the Mandelstam-Tamm relation includes the Planck constant, which determines the time scale of quantum systems. 
In the classical case, one cannot use the Planck constant anymore, but one still anticipates some classical version of speed limit. 
Quite recently, there have been several attempts of formal extension to closed classical Hamiltonian systems~\cite{OO, Campo}. 
In classical stochastic systems, there exist arguments on classical overdamped Langevin systems~\cite{Sekimoto-Sasa, Aurell}. 
However, the proofs strongly rely on the speciality of the overdamped Langevin equation, and their extension to general stochastic systems does not look easy. 
As a general consensus for classical stochastic processes, it is still an open problem to figure out general physical quantities playing similar roles to energy fluctuation and the Planck constant, which are key-ingredients in quantum speed limit.

In this Letter, we put forward this direction considering general stochastic processes with discrete states. 
We employ the techniques developed in Ref.~\cite{SST}, and derive speed limit inequalities for general classical stochastic processes. 
In this inequality, the operation time is bounded below by the combination of the entropy production~\cite{Sei} and the dynamical activity~\cite{Maes1, Maes2}. 
Since this bound is not tight in systems with nonzero stationary current, we further generalize this inequality by introducing the Hatano-Sasa entropy production~\cite{HS}. 
Here, the entropy production quantifies the irreversibility, and the dynamical activity characterizes the time scale in the nonequilibrium dynamics of the system. 
The Hatano-Sasa entropy production is an extension of the entropy production to systems with stationary current. 
Thus, our inequalities clarify the trade-off between the degree of irreversibility and the speed of state transformation in classical stochastic systems. 
In addition, they figure out clear physical pictures which have very similar structure to the quantum speed limit such as Mandelstam-Tamm and Margolus-Levitin relations.

\para{Setup and first main result}
Consider a classical stochastic Markov process with discrete states. 
Let $p_i(t)$ and $W_{ij}(t)$ be the probability distribution of the state $i$ and the transition matrix element of the transition $j\to i$ at time $t$. 
The time evolution of the probability distributions is given by the following master equation:
\eq{
\frac{d}{dt}p_i(t)=\sum_j W_{ij}(t)p_j(t).
}
The transition matrix satisfies the normalization condition $\sum_i W_{ij}(t)=0$ and nonnegativity $W_{ij}(t)\geq 0$ for $i\neq j$.
We consider the state transformation from ${\bm p}(0)$ to ${\bm p}(\tau)$ by changing the transition matrix in $0\leq t\leq \tau$.
We measure the distance of two probability distributions ${\bm p}$ and ${\bm p}'$ by the so-called statistical distance, or variation distance~\cite{CT}, defined as
\eq{
L({\bm p},{\bm p}'):=\sum_i \abs{p_i-p'_i}.
}

We first consider systems with time reversal symmetry (i.e., the local detailed-balance condition); $W_{ij}(t)e^{-\beta E_{j}(t)}=W_{ji}(t)e^{-\beta E_{i}(t)}$. 
Here, $E_{i}(t)$ is the energy of the $i$'th state of the system and $\beta$ is the inverse temperature of the heat bath inducing this transition.
The heat absorption by the bath associated with the transition $i\to j$ is given by $Q_{i\to j}(t)=E_i(t)-E_j(t)$.
If the system is attached to multiple reservoirs, the local detailed-balance condition and heat absorption are considered for each reservoir.
We now introduce the entropy production defined as the sum of the entropy increase of the heat bath and the difference in the Shannon entropy of the system:
\eq{
S(t):=-\sum_i p_i(t)\ln p_i(t).
}
The entropy production rate $\dsgm$ and the total entropy production $\Sigma$ are thus given by~\cite{Sei}
\balign{
\dsgm (t):=&\frac{d}{dt}S(t)+\sum_{i\neq j} W_{ji}(t)p_i(t) \beta Q_{i\to j} (t), \\
\Sigma:=&\int_0^\tau dt \dsgm(t).
}
The entropy production quantifies the irreversibility of a thermodynamic process. 
In fact, we can show the second law of thermodynamics by using the local detailed-balance condition and the nonnegativity of the relative entropy as follows:
\eq{
\dsgm (t)=\sum_{i\neq j}W_{ji}(t)p_i(t) \ln \frac{W_{ji}(t)p_i(t)}{W_{ij}(t)p_j(t)} \geq 0.
}
The equality holds only when the probability distribution ${\bm p}(t)$ is the instantaneous canonical distribution for the transition matrix ${\bm W}(t)$.

To quantify the system's inherent time scale, we employ the dynamical activity
\eq{
A(t):=\sum_{i \neq j} {W_{ij}(t)p_j(t)}.
}
The dynamical activity is originally defined through the time-symmetric part in the action in the stochastic path probability (c.f., the time-asymmetric part is the entropy production)\cite{Maes1, Maes2}, and is considered to play a key role in nonequilibrium dynamics \cite{maesbook}.
The dynamical activity quantifies how frequently a state changes, and thus it characterizes the time scale of the system.
The time average of the activity is denoted by $\la A\ra _\tau :=\frac{1}{\tau}\int_0^\tau dt A(t)$.
We now claim the following speed-limit inequality
\eqa{
\tau_{\rm I}:=\frac{L({\bm p}(0),{\bm p}(\tau))^2}{2\Sigma \la A\ra_\tau}
\leq \tau .
}{first-main}
This is our first main result.
This inequality clearly shows that the product of the entropy production and the time-averaged activity bounds the speed of the state transformation.
Remarkably, the relation (\ref{first-main}) has a similar structure to the typical expression of quantum speed limit, where the dynamical activity plays a similar role to the Planck constant in the quantum speed limit.

\para{Second main result}
Although the inequality \eqref{first-main} provides a good bound in systems in equilibrium environment, it becomes a weak bound in systems attached to multiple reservoirs with different temperatures, and in particular this inequality is trivial in the limit of large $\tau$ (i.e., $\tau_I \to 0$). 
This is because the entropy production increases linearly in time due to the stationary heat current.
Thus, in general cases with finite stationary currents, we need to derive another type of relation to obtain a useful bound on operation time.
For this aim, we distinguish the entropy production caused by the state transformation and that by the stationary dissipation. 
Assuming the uniqueness of the stationary state, we introduce {\it the generalized heat} (or excess heat) defined as
\eq{
\beta \qex _{i\to j}(t):=\ln \frac{\pss _j(t)}{\pss _i(t)}.
}
Here, ${\bm p}^{\rm ss} (t)$ is the instantaneous stationary probability distribution for the transition matrix ${\bm W}(t)$.
If the stationary state is an equilibrium state, this quantity reduces to the usual expression of the entropy production in the heat bath.
Using this, we define the Hatano-Sasa entropy production rate and the total Hatano-Sasa entropy production as~\cite{HS, dif-HS}
\balign{
\dsgmhs (t):=&\frac{d}{dt}S(t)+\sum_{i,j}W_{ji}(t)p_i(t) \beta \qex _{i\to j}(t), \\
\Sigma^{\rm HS}:=&\int_0^\tau dt \dsgmhs (t).
}
Notably, even when finite stationary currents exist, the Hatano-Sasa entropy production vanishes for a quasistatic process \cite{HS}. In addition, the fluctuation theorem can be formulated by using the stochastic counterpart of the Hatano-Sasa entropy production.
Therefore, the Hatano-Sasa entropy production is considered to capture the thermodynamic properties of general stochastic processes.
Using the Hatano-Sasa entropy production, we have another speed-limit inequality
\eqa{
\tau_{\rm II}:=\frac{c^* L({\bm p}(0),{\bm p}(\tau))^2}{2\Sigma^{\rm HS} \la A\ra_\tau}\leq \tau .
}{second-main}
This is our second main result.
Here, $c^*:=0.896\cdots$ is a solution of
\eq{
c^*=\min_y \frac{(1-e^y+ye^y)(1+e^y)}{(1-e^y)^2}.
}
The inequality \eqref{second-main} shows that the Hatano-Sasa entropy production is a key-ingredient to characterize a bound on the operation time, when the system has a finite stationary current.
In addition, the dynamical activity again plays a crucial role to determine the time scale. 
We remark that the generalized heat works in general stochastic systems without the local detailed-balance condition, and so as our second main result.

\para{Example: two-level system}
Before going to the proofs, we demonstrate our inequalities~\eqref{first-main} and \eqref{second-main} with a simple solvable model. 
This system consists of two states, 0 and 1, whose energies are given by $E_0=0$ and $E_1(t)$, respectively. Suppose that the initial distribution is $p_0(0)=p_1(0)=1/2$, and we transform them to $p_0(\tau)=3/4$ and $p_1(\tau)=1/4$ with the time interval $\tau$. As we see below, even such a simple model is very instructive to understand the physical structure of our results.

\figin{8.8cm}{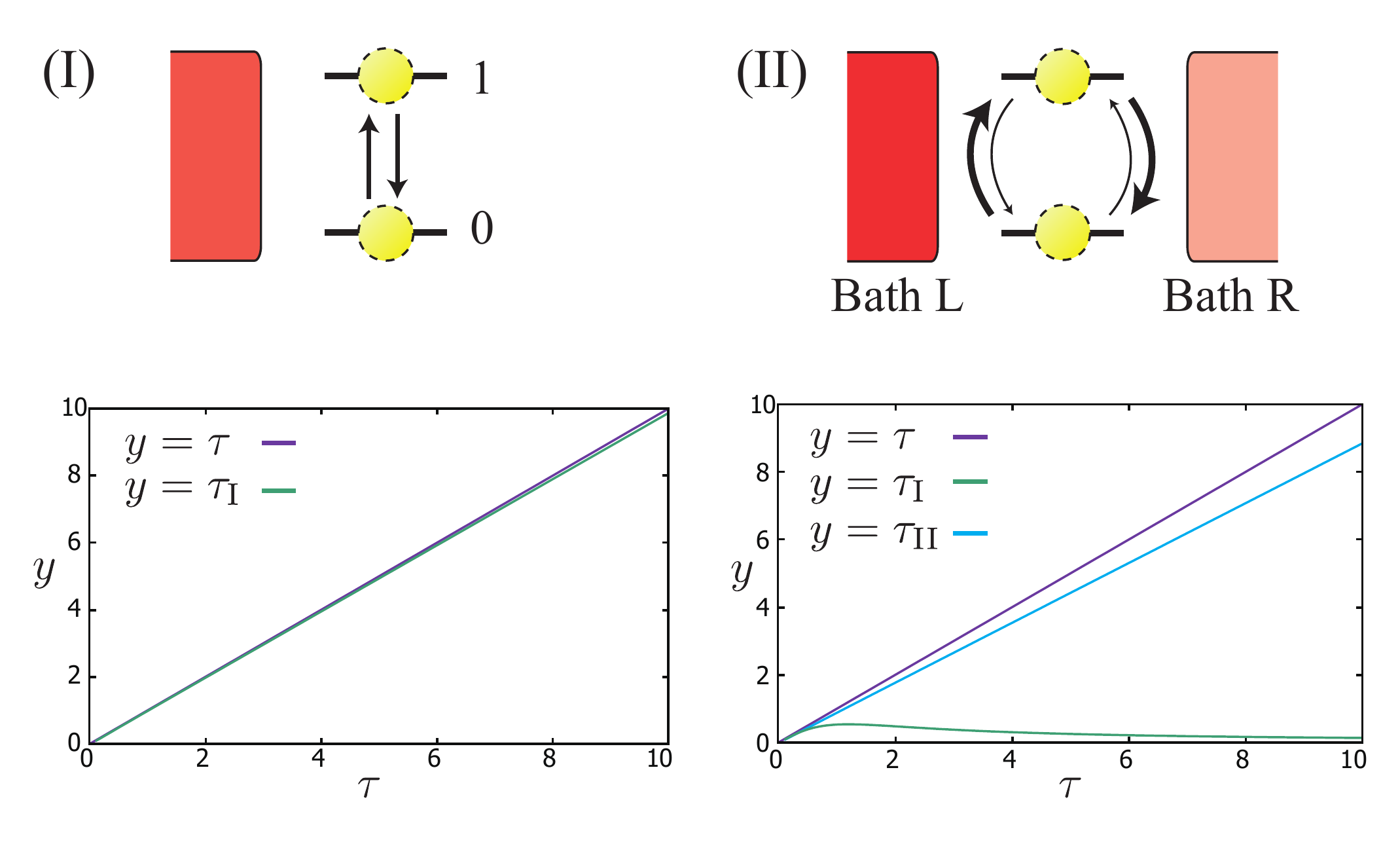}{
(I): Demonstration of the relation (\ref{first-main}) in case with a single bath.
In the bottom plot, the purple line is the linear reference line $y=\tau $, and the green curve is $y=\tau_{\rm I}$ as a function of $\tau$. 
These two lines almost agree, which clearly shows the tightness of the bound (\ref{first-main}).
(II): Demonstration of the relation (\ref{second-main}) in case with two baths with $\alpha=2/3$ (finite stationary current exists).
The bottom plot shows that \eref{first-main} ($\tau_{\rm I}\leq \tau$) is a poor bound, while \eref{second-main} ($\tau_{\rm II}\leq \tau$) still provides a meaningful bound.
}{q-dot}

We first consider the case with a single heat bath with inverse temperature $\beta$ (See \fref{q-dot}.(I)).
For convenience of calculation, we set the transition matrix as
\eq{
W_{10}=1, \ \ \ \ W_{01}=e^{\beta E_1(t)} \nt
}
with
\eq{
E_1(t):=\frac1\beta \ln \( \frac{4\tau +1}{2\tau -t} -1\) ,
}
which provides the solution $p_1(t)=\frac12 -\frac{t}{4\tau }$.
Then, it is straightforward to get the dynamical activity and the distance as $A(t)=1+\frac{1}{4\tau }+\frac{t}{2\tau }$ and $L(\bsp(0),\bsp(\tau))=\frac12$.
The averaged activity $\la A\ra_\tau =\frac52 +\frac1{2\tau }$ is a quantity of $O(1)$.
The bound $\tau_{\rm I}$ on the operation time is explicitly given by
\eqa{
\tau_{\rm I}=
\frac{1}{(10+2/\tau )\Sigma} \, ,
}{toy-first}
where the entropy production $\Sigma$ is given through a little complicated calculation:
\eqa{
\Sigma =\frac{1}{4\tau }\int_{2\tau }^{2\tau +1} dy \ln \( 1+\frac{\tau }{y}\) .
}{toy-Sigma}
From these expressions, one can directly confirm the validity of (\ref{first-main})~\cite{direct}. 
The asymptotic behavior in large $\tau$ reads $\frac{1}{10\Sigma} \leq \tau $, which is a very good estimation since $\Sigma \simeq \frac1{4\tau }\ln \frac32=0.101\cdots \times \frac1\tau $. We show the plot of these results in \fref{q-dot}.(I).

We next consider the case with two heat baths, L and R to demonstrate the relation (\ref{second-main}) (See \fref{q-dot}.(II)). The transition matrices associated with each bath are set as
\balign{
W^{\rm L}_{10}=&\alpha , \ &W^{\rm L}_{01}=\frac12 \( \frac{4\tau +1}{2\tau -t} -1\) , \nt \\
W^{\rm R}_{10}=&1-\alpha , \ &W^{\rm R}_{01}=\frac12 \( \frac{4\tau +1}{2\tau -t} -1\) , \nt
}
with $\alpha \neq \frac12$, which ensures the existence of nonzero stationary current between L and R in the stationary state.
The probability distribution and the dynamical activity are completely the same as in the single-bath case; $p_1(t)=\frac12 -\frac{t}{4\tau }$ and $A(t)=1+\frac{1}{4\tau }+\frac{t}{2\tau }$.
In contrast to these quantities, the entropy production is larger than that in the single-bath case (The explicit form of entropy production is presented in \cite{noneq-sigma}).
In particular, for large $\tau$, the entropy production asymptotically behaves as $\Sigma \simeq \frac{5\tau }8 \( \frac12 -\alpha \) \ln \frac{1-\alpha}\alpha $, which increases in proportion to $\tau $.
Thus, the first inequality \eqref{first-main} falls into a trivial bound in this case (i.e., $\tau_{\rm I}\to 0$ in $\tau \to \infty$ limit).

On the other hand, the Hatano-Sasa entropy production is given by
\eq{
\Sigma^{\rm HS} =\frac{1}{4\tau }\int_{2\tau }^{2\tau +1} dy \ln \( 1+\frac{\tau }{y}\) ,
}
which is exactly same as the entropy production \eqref{toy-Sigma} in the single-bath case.
Hence the Hatano-Sasa entropy production is finite for large $\tau$. 
The bound $\tau_{II}$ on the operation time is given by
\eqa{
\tau_{II}= 
\frac{c^*}{(10+2/\tau )\Sigma^{\rm HS}} ,
}{toy-second}
which provides a meaningful bound even for large $\tau $; an asymptotic bound $c^*/(10\Sigma^{\rm HS})\leq \tau $. See the plot in \fref{q-dot}.(II).

\para{Proof of the first main result \eqref{first-main}}
The crucial relation in this proof is the following inequality for instant quantities:
\eqa{
\sum_i \abs{\frac{d}{dt} p_i(t)}\leq \sqrt{2\dsgm(t) A(t)}.
}{first-key}
In the derivation, we drop the time dependence for visibility.

The entropy production rate is evaluated as
\balign{
\dsgm 
=&\frac{1}{2}\sum_{i\neq j} (W_{ji}p_i-W_{ij}p_j) \ln \frac{W_{ji}p_i}{W_{ij}p_j} \nt \\
\geq&\sum_{i\neq j}  \frac{(W_{ji}p_i-W_{ij}p_j)^2}{W_{ji}p_i+W_{ij}p_j},
}
which follows from a simple inequality $(a-b)\ln (a/b)\geq 2(a-b)^2/(a+b)$.
Using the master equation and the Schwarz inequality, we obtain the relation \eqref{first-key}:
\balign{
&\sum_i \abs{\frac{d}{dt} p_i}=\sum_i \abs{\sum_{j(\neq i)} W_{ij}p_j-W_{ji}p_i} \nt \\
\leq&\sqrt{\( \sum_{i \neq j} \frac{(W_{ij}p_j-W_{ji}p_i)^2}{W_{ij}p_j+W_{ji}p_i}\) \( \sum_{i \neq j} {W_{ij}p_j+W_{ji}p_i}\) } \nt \\
=&\sqrt{2\dsgm A}.
}
The key inequality \eqref{first-key} directly implies the desired result, \eref{first-main}:
\balign{
\sum_i \abs{p_i(0)-p_i(\tau)} 
\leq&\sum_i \int_0^\tau dt \abs{\frac{d}{dt} p_i} \nt \\
\leq&\int_0^\tau dt \sqrt{2\dsgm A} 
\leq\sqrt{2\Sigma \tau \la A\ra _\tau}. \lb{last-trans}
}
At the last inequality, we used the Schwarz inequality. It is also straightforward to show the result for multiple reservoirs by starting with the expression of entropy production rate caused by each reservoir.

\para{Proof of the second main result \eqref{second-main}}
The derivation of \eref{second-main} is similar to that of \eref{first-main}.
The crucial relation in this derivation is
\eqa{
\sum_i \abs{\frac{d}{dt} p_i(t)}\leq \sqrt{\dsgmhs (t) \frac{2A(t)}{c^*}}.
}{second-key}
We again drop the time dependence for visibility.

To derive \eref{second-key}, we introduce the dual transition matrix $\tw$~\cite{Kam} given by
\eq{
\tw _{ij}:=\frac{W_{ji}\pss _i}{\pss _j}.
}
We can easily check that $\tw$ is indeed a transition matrix (i.e., $\tw$ satisfies the normalization condition and nonnegativity).
It is noteworthy that the diagonal element of the dual transition matrix and the original one are equal:
\eqa{
\sum_{j(\neq i)} \tw_{ji}=-\tw_{ii}=-W_{ii}=\sum_{j(\neq i)} W_{ji}.
}{dual-diag}
The dual transition matrix provides another expression of the Hatano-Sasa entropy production rate
\balign{
\dsgmhs &=\sum_{i\neq j}W_{ji}p_i \ln \frac{W_{ji}p_i}{\tw_{ij}p_j} \nt \\
&=\sum_{i\neq j}W_{ji}p_i \ln \frac{W_{ji}p_i}{\tw_{ij}p_j}+\tw_{ij}p_j-W_{ji}p_i, \nt \\
\intertext{where we used \eref{dual-diag}.
This is a generalized form of the partial entropy production~\cite{SS15, SS}.
Using an inequality $a\ln a/b +b-a\geq c^*(a-b)^2/(a+b)$ for nonnegative $a$ and $b$~\cite{SST}, the right-hand side is bounded below by
}
&\geq c^* \sum_{i\neq j}  \frac{(W_{ji}p_i-\tw_{ij}p_j)^2}{W_{ji}p_i+\tw_{ij}p_j}.
}
We then arrive at the key relation \eqref{second-key}
\balign{
&\sum_i \abs{\frac{d}{dt} p_i}=\sum_i \abs{\sum_{j(\neq i)} W_{ij}p_j-\tw_{ji}p_i} \nt \\
\leq&\sqrt{\( c^* \sum_{i \neq j} \frac{(W_{ij}p_j-\tw_{ji}p_i)^2}{W_{ij}p_j+\tw_{ji}p_i}\) \( \frac{1}{c^*} \sum_{i \neq j} (W_{ij}p_j+W_{ji}p_i) \) } \nt \\
\leq&\sqrt{\dsgmhs  \, \frac{2A}{c^*}}.
}
In the second line, we used \eref{dual-diag}.
Following the same procedure as in \eref{last-trans}, \eref{second-key} leads to the desired inequality \eqref{second-main}.

\para{Discussion}
We have established fundamental trade-off inequalities in general stochastic processes which claim that increasing the speed of the state transformation inevitably requires large entropy production or large Hatano-Sasa entropy production. 
This shows clear contrast to the case of an isolated quantum system, where the energy fluctuation plays a role to bound the speed of the state transformation. 
The coefficient appearing in these inequalities is the dynamical activity, which determines the time scale and thus it plays a similar role to the Planck constant in quantum version of speed limit.
The Hatano-Sasa entropy production was first introduced to construct an extended Clausius inequality and a framework of steady state thermodynamics~\cite{HS}.
Our result elucidates the role of the Hatano-Sasa entropy production from a new point of view.

Probabilistic systems with discrete states are ubiquitous in nature, such as proteins in biosystems~\cite{Sek, cell} and quantum dots in the classical regime~\cite{Fuj, Kun}, name only a few. 
Thus, our inequalities can be experimentally tested at the quantitative level. 
Especially quantum-dot systems have high-controllability with high accuracy nowadays ~\cite{Fuj, Kun, Pek, Pek2, Pek3}, and is the most promising experimental object.

\para{Acknowledgement}
NS was supported by Grant-in-Aid for JSPS Fellows JP17J00393. 
KF acknowledges supports from the National Science Foundation of China under Grants No.~11375012 and 11534002, and The Recruitment Program of Global Youth Experts of China.
KS was supported by JSPS Grants-in-Aid for Scientific Research (No. JP25103003, JP16H02211 and JP17K05587).


\begin{thebibliography}{99}

\bib{MT}
L. Mandelstam and I. Tamm, {\em The uncertainty relation between energy and time in nonrelativistic quantum mechanics}. J. Phys. (USSR) {\bf 9}, 249 (1945).
\bib{Fle}
G. N. Fleming, {\em A unitarity bound on the evolution of nonstationary states}. Nuovo Cimento A {\bf 16}, 232 (1973).
\bib{AA}
J. Anandan and Y. Aharonov, {\em Geometry of quantum evolution}. Phys. Rev. Lett. {\bf 65}, 1697 (1990).
\bib{Pre}
P. Pfeifer, {\em How fast can a quantum state change with time?}. Phys. Rev. Lett. {\bf 70}, 3365 (1993).
\bib{ML}
N. Margolus and L. B. Levitin, {\em The maximum speed of dynamical evolution}. Physica D {\bf 120}, 188 (1998).
\bib{GLM}
V. Giovannetti, S. Lloyd, and L. Maccone, {\em Quantum limits to dynamical evolution}. Phys. Rev. A 67, 052109 (2003).
\bib{JK}
P. J. Jones and P. Kok, {\em Geometric derivation of the quantum speed limit}. Phys. Rev. A {\bf 82}, 022107 (2010).

\bib{DC}
S. Deffner and S. Campbell, {\em Quantum speed limits: from Heisenberg's uncertainty principle to optimal quantum control}. J. Phys. A: Math. Theor. {\bf 50,} 453001 (2017).

\bib{Funo}
K. Funo, J.-N. Zhang, C. Chatou, K. Kim, M. Ueda, and A. del Campo, {\em Universal Work Fluctuations During Shortcuts to Adiabaticity by Counterdiabatic Driving}. Phys. Rev. Lett. {\bf 118}, 100602 (2017).

\bib{Tad}
M. M. Taddei, B. M. Escher, L. Davidovich, and R. L. de Matos Filho, {\em Quantum Speed Limit for Physical Processes}. Phys. Rev. Lett. {\bf 110}, 050402 (2013).
\bib{Cam}
A. del Campo, I. L. Egusquiza, M. B. Plenio, and S. F. Huelga, {\em Quantum Speed Limits in Open System Dynamics}. Phys. Rev. Lett. {\bf 110}, 050403 (2013).
\bib{DL}
S. Deffner and E. Lutz, {\em Quantum Speed Limit for Non-Markovian Dynamics}. Phys. Rev. Lett. {\bf 111}, 010402 (2013).
\bib{Zha}
Y.-J. Zhang, W. Han, Y.-J. Xia, J.-P. Cao, and H. Fan, {\em Quantum speed limit for arbitrary initial states}. Sci. Rep. {\bf 4}, 4890 (2014).
\bib{Pir} D. P. Pires, M. Cianciaruso, L. C. C\'{e}leri, G. Adesso, and D. O. Soares-Pinto, {\em Generalized Geometric Quantum Speed Limits}. Phys. Rev. X {\bf 6,} 021031 (2016).


\bib{OO}
M. Okuyama and M. Ohzeki, {\em Quantum Speed Limit is Not Quantum}. Phys. Rev. Lett. {\bf 120}, 070402 (2018).
\bib{Campo}
B. Shanahan, A. Chenu, N. Margolus, A. del Campo, {\em Quantum Speed Limits Across the Quantum-to-Classical Transition}. Phys. Rev. Lett. {\bf 120}, 070401 (2018).


\bibitem{Sekimoto-Sasa}
K. Sekimoto and S.-i. Sasa, {\em Complementarity relation for irreversible process derived from stochastic energetics}. J. Phys. Soc. Jpn. {\bf 66}, 3326 (1997).
\bibitem{Aurell}
E. Aurell, K. Gaw\c{e}dzki , C. Mej\'{\i}a-Monasterio, R. Mohayaee, P. Muratore-Ginanneschi, {\em Refined second law of thermodynamics for fast random processes}. J. Stat. Phys. {\bf 147}, 487 (2012).


\bib{SST}
N. Shiraishi, K. Saito, and H. Tasaki, {\em Universal Trade-Off Relation between Power and Efficiency for Heat Engines}. Phys. Rev. Lett. {\bf 117}, 190601 (2016).


\bibitem{Sei} 
U. Seifert, {\em Stochastic thermodynamics, fluctuation theorems, and molecular machines}. Rep. Prog. Phys. {\bf 75}, 126001 (2012).


\bib{Maes1}
M. Baiesi, C. Maes, and B. Wynants, {\em Fluctuations and Response of Nonequilibrium States}. Phys. Rev. Lett. {\bf 103}, 010602 (2009).
\bib{Maes2}
M. Baiesi, C. Maes, and B. Wynants, {\em Nonequilibrium Linear Response for Markov Dynamics, I: Jump Processes and Overdamped Diffusions}. J. Stat. Phys. {\bf 137}, 1094 (2009).

\bib{HS}
T. Hatano and S.-i. Sasa, {\em Steady-State Thermodynamics of Langevin Systems}. Phys. Rev. Lett. {\bf 86}, 3463 (2001).

\bib{CT}
T. M. Cover and J. A. Thomas, {\em Elements of Information Theory}, 2nd ed. Wiley (2006).

\bib{maesbook}
C. Maes, {\em Non-Dissipative Effects in Nonequilibrium Systems}, Springer (2018).

\bib{dif-HS}
We remark that unlike the original Hatano-Sasa paper, the change in entropy is formulated as in the form of Shannon entropy of the present distribution.





\bib{direct}
The inequality \eqref{first-main} is directly checked by using $\Sigma >\ln \frac32 -\frac{\tau }{2(2\tau +1)(3\tau +1)}$ and $\ln \frac32 >\frac25 +\frac1{200}$.


\bib{noneq-sigma}
The explicit form of the entropy production is
\balign{
\Sigma = &\frac{1}{4\tau }\int_{2\tau }^{2\tau +1} dy \ln \( 1+\frac{\tau }{y}\) +\frac18 \ln (4\alpha (1-\alpha )) \nt \\
&+ \frac{5\tau }8 \( \frac12 -\alpha \) \ln \frac{1-\alpha}\alpha ,
}
where the second and third terms are newly appears compared to the single-bath case.

\bib{Kam}
N. G. Van Kampen, {\em Stochastic Process in Physics and Chemistry}, 3rd ed. Elsevier (2007).




\bib{SS15}
N. Shiraishi and T. Sagawa, {\em Fluctuation theorem for partially masked nonequilibrium dynamics}. Phys. Rev. E {\bf 91}, 012130 (2015).
\bib{SS}
N. Shiraishi and K. Saito, {\em Incompatibility between Carnot efficiency and finite power in Markovian dynamics}. arXiv: 1602.03645 (2016).

\bib{Sek}
K. Sekimoto, {\em Stochastic energetics}. Springer (2010).
\bib{cell}
R. Phillips, J. Kondev, J. Theriot, and H. Garcia, {\it Physical Biology of the Cell}, 2nd ed. Garland Science (2012).


\bib{Fuj}
T. Fujisawa, T. Hayashi, R. Tomita, and Y. Hirayama, {\em Bidirectional counting of single electrons}. Science {\bf 312}, 1634 (2006).
\bib{Kun}
B. K\"{u}ng, C. R\"{o}ssler, M. Beck, M. Marthaler, D. S. Golubev, Y. Utsumi, T. Ihn, and K. Ensslin, {\em Irreversibility on the Level of Single-Electron Tunneling}. Phys. Rev. X {\bf 2}, 011001 (2012).

\bib{Pek}
J. V. Koski, V. F. Maisi, J. P. Pekola and D. V. Averin, {\em Experimental realization of a Szilard engine with a single electron}. Pro. Nat. Ac. Sci. {\bf 111}, 13786 (2012).
\bib{Pek2}
J. V. Koski, V.F. Maisi, T. Sagawa, and J. P. Pekola, {\em Experimental Observation of the Role of Mutual Information in the Nonequilibrium Dynamics of a Maxwell Demon}. Phys. Rev. Lett. {\bf 113}, 030601 (2014).
\bib{Pek3}
J. V. Koski, A. Kutvonen, I. M. Khaymovich, T. Ala-Nissila, and J. P. Pekola, {\em On-Chip Maxwell’s Demon as an Information-Powered Refrigerator}. Phys. Rev. Lett. {\bf 115}, 260602 (2015).






\end{thebibliography}
\end{document}